\documentclass[sigconf,nonacm]{acmart}
\usepackage{amsmath,amsfonts}
\usepackage{subcaption}
\usepackage{float}
\usepackage{algorithm}
\usepackage[noend]{algpseudocode}
\usepackage[frozencache,cachedir=.]{minted}
\usepackage{array}
\usepackage{booktabs}
\usepackage{longtable}
\usepackage{caption}
\usepackage{tablefootnote}
\usepackage{multirow}

\usepackage{enumitem,listings}
\usepackage[normalem]{ulem}
\usepackage[T1]{fontenc}
\usepackage[scaled=0.85]{beramono}

\newif\ifshort
\shortfalse
\ifshort
\newcommand*{\RefAlgTitle}[1]{\lstinline{#1}}
\else
\newcommand*{\RefAlgTitle}[1]{\hyperref[alg:#1]{\lstinline{#1}}}
\fi

\newcommand*{\RefS}[1]{\hyperref[#1]{\S\ref*{#1}}}
\newcommand*{\RefSection}[1]{\hyperref[#1]{Section~\ref*{#1}}}
\newcommand*{\RefAppendix}[1]{\hyperref[#1]{Appendix \ref*{#1}}}
\newcommand*{\RefRemark}[1]{\hyperref[remark:#1]{Remark #1}}
\newcommand*{\RefInvar}[1]{\hyperref[#1]{Invariant \ref*{#1}}}
\newcommand*{\RefLemma}[1]{\hyperref[lemma:#1]{Lemma #1}}
\newcommand*{\RefAlgorithm}[1]{\hyperref[#1]{Algorithm \ref*{#1}}}

\lstdefinelanguage{Go2}{
    morekeywords=[1]{break,default,func,interface,select,case,defer,go,map,vid,struct,chan,else,goto,package,switch,const,fallthrough,if,range,in,type,continue,for,import,return,var,while,and,or,xor,not},
    morekeywords=[2]{append,cap,close,complex,copy,delete,imag,len,make,new,panic,print,println,real,recover,assert},
    morekeywords=[3]{bool,byte,complex64,complex128,error,float,float32,float64,int,int8,int16,int32,int64,rune,string,uint,uint8,uint16,uint32,uint64,uintptr},
    morekeywords=[4]{true,false,iota,nil},
    morestring=[b]{"},
    morestring=[b]{'},
    morestring=[b]{`},
    comment=[l]{//},
    morecomment=[s]{/*}{*/},
    sensitive=true
}

\DeclareCaptionFormat{lstcaptionformat}{\small\textit{#1#2#3}\par\vspace{1pt}\hrule}
\DeclareCaptionFormat{lstcaptionformatnoalg}{\small\textit{#3}\par\vspace{1pt}\hrule}
\setcopyright{none}
\renewcommand\footnotetextcopyrightpermission[1]{}
\makeatletter
\def\@copyrightpermission{}
\makeatother
\thanks{Preprint}

\begin{document}

\title{\textit{SSSP-Del}: Fully Dynamic Distributed Algorithm for Single-Source Shortest Path}

\author{Parshan Javanrood}
\email{pjavan@student.ubc.ca}
\orcid{0009-0006-5293-1108}
\affiliation{%
  \institution{University of British Columbia}
  \city{Vancouver}
  \state{BC}
  \country{Canada}
}

\author{Matei Ripeanu}
\email{matei@ece.ubc.ca}
\orcid{https://orcid.org/0000-0001-9839-3866}
\affiliation{%
  \institution{University of British Columbia}
  \city{Vancouver}
  \state{BC}
  \country{Canada}
}

\renewcommand{\shortauthors}{Javanrood et al.}

\begin{abstract}
Modern graphs are both large and dynamic, presenting significant challenges for fundamental queries, such as the Single-Source Shortest Path (SSSP) problem. Naively recomputing the SSSP tree after each topology change is prohibitively expensive, causing on-demand computation to suffer from high latency. Existing dynamic SSSP algorithms often cannot simultaneously handle both edge additions and deletions, operate in distributed memory, and provide low-latency query results. To address these challenges, this paper presents SSSP-Del, a new vertex-centric, asynchronous, and fully distributed algorithm for dynamic SSSP. Operating in a shared-nothing architecture, our algorithm processes streams of both edge insertions and deletions. We conduct a comprehensive evaluation on large real-world and synthetic graphs with millions of vertices, and provide a thorough analysis by evaluating result latency, solution stability, and throughput.
\end{abstract}


\maketitle

\section{Introduction\label{sec:introduction}}
Modern graphs are both massive and highly dynamic, driven by a continuous stream of updates \cite{besta2023streaming}. Online analytics on such \textit{streaming graphs} are critical for many applications (e.g., security, monitoring social networks, or traffic routing) and pose unique challenges \cite{besta2023streaming, li2024lsmgraph}.

The Single-Source Shortest Path (SSSP) problem is a fundamental graph query used in routing, connectivity, and network analysis. In a \textit{static} graph, classic algorithms like Dijkstra or Bellman–Ford compute the \textit{shortest-path tree}, the spanning three that is the union of all shortest paths between the source and all graph vertices. In a \textit{dynamic} graph, however, naively re-computing the shortest-path tree after each graph topology change is prohibitively expensive \cite{agarwal2016parallel}. 

Instead, \textit{dynamic} SSSP algorithms seek to update only the portion of the shortest path tree impacted by the topology change(s). Edge \textit{insertions} are relatively easy to support~\footnote{Assuming positive edge weights.}, as they result in limited restructuring of the shortest-path tree.  Edge \textit{deletions}, however, are especially problematic as they can lead to complete restructuring of the shortest-path tree \cite{riazi2018sssp}. In short, maintaining the shortest path tree under dynamic conditions requires careful update propagation and invalidation mechanisms, efficient data structures to track affected vertices, and supporting concurrency to handle high volumes of topology updates \cite{agarwal2016parallel, riazi2018sssp}.

While variety of dynamic SSSP solutions have been proposed  \cite{ramalingam1996complexity, frigioni2000fully, agarwal2016parallel, riazi2018sssp, srinivasan2018}, none of them are able to achieve all of the following goals simultaneously: \textit{(i)} handle both edge additions and deletions, \textit{(ii)} operate on distributed memory, and \textit{(iii)} provide the shortest-path three on-demand with low latency. This paper presents a new \textit{vertex-centric, shared-nothing, asynchronous} algorithm for dynamic SSSP. The key contributions are:

\begin{itemize}
    \item \textbf{A fully distributed algorithm.} We propose a vertex-centric algorithm that supports both edge insertions and deletions by propagating distance changes across a shared-nothing architecture. Designed for high-throughput topology update streams, our approach scales across distributed systems and supports low-latency, on-demand shortest-path queries.

    \item \textbf{Comprehensive evaluation.} We validate our approach on large real-world and synthetic graphs with millions of vertices and tens of millions of edges. Our experiments demonstrate scalability and responsiveness, significantly outperforming two alternatives: \textit{(i)} an optimized static graph baseline that recomputes the solution from scratch on-demand (we compare with two implementations: Galois \cite{nguyen2013lightweight, whang2015scalable} and our own ); \textit{(ii)} a stream-oriented solution (GraphBolt \cite{graphbolt2019}) that processes topology updates in batches.

\end{itemize}


\section{Problem Background\label{sec:pbackground}}

\textbf{\textit{Synchronous Algorithms on Shared-memory}}: Early work has focused on solutions to process one topology update at a time and assumed a shared memory system.  Ramalingam and Reps(RR)~\cite{ramalingam1996complexity} developed a fully dynamic SSSP algorithm based on Dijkstra’s algorithm maintains the shortest-path tree and updates it synchronously after each topology change. On edge \emph{deletion}, it proceeds in two phases: first, it identifies all \emph{affected} vertices whose shortest paths are invalidated; then it runs a Dijkstra-like update to recompute their distances. Frigioni et al.~\cite{frigioni2000fully} proposed another dynamic algorithm (FMN) that supports both insertions and deletions. The algorithm assigns each edge a \emph{forward/backward level} and an \emph{owner} endpoint, which together bound how many times the edge is processed during updates. The levels capture the difference between endpoint distances and the edge weight, and non-owned edges are kept in priority queues. On a distance update at a vertex, only edges owned by the vertex or those with a promising priority are scanned. The algorithm achieves sublinear amortized update time in the number of affected elements, but it is inherently sequential and assumes access to shared data structures.


\textbf{\textit{Distributed Dynamic SSSP}}: More recently, there has been interest in distributed dynamic SSSP frameworks. Riazi et al.~\cite{riazi2017sssp} proposed SSSPIncJoint, a distributed algorithm implemented in Apache Spark's GraphX. SSSPIncJoint models dynamic changes as a sequence of \emph{snapshots}, each after applying a batch of updates to the graph. The algorithm then incrementally updates the SSSP tree using the gather-apply-scatter (GAS) programming model. It avoids full recomputation by reusing state from previous snapshots. However, updates must be applied in bulk, and queries can only be answered at the end of each snapshot, preventing the algorithm from answering queries during intermediate update stages. 

\textbf{\textit{\textit{ReMo} Increment-Only SSSP}}: A recursive and monotonic(ReMo) algorithm for the edge-add only SSSP problem, inspired by breadth-first search (BFS), is proposed in~\cite{sallinen2021}. This method leverages the monotonicity of additive edge updates to propagate distance values efficiently without rollback. 

Traditional dynamic SSSP algorithms like RR and FMN are designed for synchronous systems and assume shared memory. Algorithms based on the batch processing model are limited in their ability to respond to queries issued at arbitrary times, and the SSSP state is only available after an entire batch is processed. Also increment-only algorithms are not fully dynamic, and do not support edge deletions.

To overcome these limitations, we propose SSSP-Del, that handles both edge additions and deletions while supporting on-demand
queries in a distributed environment. In the next section, we introduce the computational model that enables us to develop and evaluate our algorithm.


\section{Computational Model\label{sec:compmodel}}
We developed and evaluated our solution on a vertex-centric, asynchronous, shared-nothing framework. This design enables the system to (i) scale to large graphs through partitioning, (ii) respond to updates without centralized coordination, and (iii) extract results on demand with arbitrary granularity. Below, we summarize the key components of this abstraction, a more detailed formal description is provided in \cite{sallinen2023realtime}.


\textbf{Graph Representation}: Each vertex maintains a unique identifier and local state, including its in-neighbors and out-neighbors. No state is shared between vertices, enabling the graph to be partitioned across multiple compute nodes. Each node owns a disjoint subset of vertices and their edges.

\textbf{Asynchronous Vertex-Centric Execution}: Similar to HavoqGT \cite{reza2020queue,pearce2013scaling}, we adopt an asynchronous processing model in which each vertex operates as an independent agent. It reacts to two types of inputs: topology change events (e.g., edge additions or deletions) and algorithmic messages (e.g., distance updates). Upon receiving an input, the vertex executes a user-defined handler to update its local state and possibly send new messages to its neighbors. Execution proceeds in this recursive manner until no unprocessed messages or events remain. All communication is strictly one-way, FIFO-ordered, and occurs without shared memory. Topology events are prioritized over algorithmic messages to ensure structural changes are reflected promptly in downstream computations.


\textbf{Topology Event Ingestion}: The system begins with an empty graph and incrementally ingests a stream of edge-addition and edge-deletion events. These events arrive from an external source (e.g., reading from a file) with no lookahead or assumptions about future evolution. Upon receiving an event, the system updates the internal graph store and triggers the appropriate handler at the affected vertex. To delete a vertex, all its incident edges must first be removed to avoid dangling references. The system supports enforcing a lightweight barrier, referred to as an \textit{epoch}. An epoch is a convergence point in which the system temporarily pauses ingesting new topological events until all in-flight topological events and algorithmic messages have been fully handled. This guarantees that the algorithm processes all prior events before moving forward.


\textbf{Vertex-Centric Programming Interface}: Users express algorithms by defining three event handlers for messages or topology change events received at a vertex. The interface allows the user to define custom data types for vertices, edges, and messages. A vertex reacts to events and messages as described below:
\begin{itemize}
    \item \textit{onMessageReceived}: Executed when a vertex receives an algorithmic message (e.g., a distance update). The handler updates local state and may propagate messages.
    \item \textit{onEdgeChanged}: Triggered after an outgoing edge is added, deleted, or updated. 
    \item \textit{onVertexChanged}: Invoked when the vertex itself is created or deleted (excluding its edges). Typically used for initialization or cleanup.
\end{itemize}
Each handler can access and modify local vertex state, inspect metadata associated with edges or messages, and emit new messages.

\textbf{State Collection}: Results, that is, the shortest-path tree, are collected on-demand triggered by the detection of a marker event in the input stream. Before presenting the result, the framework enforces an epoch, allowing ongoing computations to complete by pausing ingestion of topological events while completing the processing of all algorithmic updates until convergence.

\section{SSSP with Deletion for Dynamic Graphs\label{sec:algo}}
\subsection{Algorithm Overview}
To ensure correctness(as we prove in \RefAppendix{app:proof}), edge deletions are handled separately from edge additions. The system continuously ingests edge addition events and propagates distance updates until it encounters a deletion. Upon detecting a deletion, it pauses further ingestion and waits for the algorithm to converge—i.e., for all in-flight messages from previous additions to be fully processed. \\
Once convergence is confirmed, the system ingests the edge deletion and again waits for the algorithm to converge in deletion mode, which operates in two distinct phases:
\begin{itemize}
    \item Invalidation Phase: All vertices whose shortest paths included the deleted edge are marked as affected. This effectively invalidates the impacted portion of the shortest-path tree.
    \item Recomputation Phase: Each invalidated vertex queries its incoming neighbors for the shortest alternative distance they can offer. These neighbors respond with their best-known distances, and the vertex selects the shortest among them to update its path.
\end{itemize}
After all deletion-related messages are processed and the system reaches convergence, the infrastructure resumes normal ingestion of topological events.

\subsection{Vertex Properties}
At vertex level, we preserve the following information:
\begin{itemize}
    \item \texttt{Distance} between the source and vertex.
    \item \texttt{PredecessorVertex} is the predecessor of the vertex in the shortest-path tree.
    \item \texttt{IncomingVertices} is the set of vertices that have an edge into the vertex.
    \item \texttt{SuccessorVertices} is the set of successors of this vertex in the shortest-path tree
    \item \texttt{MarkedAsInfinity} is a boolean flag that shows this vertex was recently impacted by an edge deletion
\end{itemize}
\begin{lstlisting}[caption=Vertex Properties]
	Distance             float64
	PredecessorVertex    *Vertex
	SuccessorVertices    map[uint32]bool // Set
	MarkedAsInfinity     bool
\end{lstlisting}

\subsection{Message Types}
Vertices communicate using messages, and below are the different types of messages:
\begin{itemize}
\item \texttt{DistanceQuery}: Sent by vertex $u$ to vertex $v$ to request the length of the shortest path from the source to $u$ through $v$
\item \texttt{DistanceUpdate}: Sent by vertex $u$ to vertex $v$ to inform $v$ of a shortest path from the source through $u$. 
\item \texttt{SetToInfinity}: Received by vertex $v$ to indicate that a previously known shortest path is no longer valid due to an edge deletion. 
\item \texttt{AddToSuccessor}: Sent by vertex $u$ to vertex $v$ to notify $v$ that $u$ is now a successor of $v$ in the shortest-path tree.
\item \texttt{RemoveFromSuccessor}: Sent by vertex $u$ to vertex $v$ to notify $v$ that $u$ is no longer a successor of $v$ in the shortest-path tree.
\end{itemize}

\begin{lstlisting}[caption=Message Types]
	DistanceQuery          // Body: {}
	DistanceUpdate         // Body: {Distance}
	SetToInfinity          // Body: {}
	AddToSuccessor         // Body: {}
	RemoveFromSuccessor    // Body: {}
\end{lstlisting}

\subsection{Topological Events}
\textbf{Edge Addition}:
When an edge addition is detected(from $u$ to $v$), the tail of the edge($u$) will inform the head($v$) of the newly computed shortest path from the source. The update effectively flows downward to update the newly connected vertices to, if possible, reduce their distance from the source.
\begin{lstlisting}[caption=Edge Addition]
func onEdgeAddition(u *Vertex, newEdge *Edge):
    sendMessage(
        edge.Destination,           // To
        DistanceUpdate,             // Type
        u.Distance + newEdge.Weight // Body
    )
\end{lstlisting}
This follows the same principle as ReMo Increment-Only SSSP~\cite{sallinen2021}, where edge additions propagate shorter distances through the graph. Due to monotonicity, distances only decrease, and vertices can process these updates asynchronously, without relying on message ordering~\cite{sallinen2019incremental}. We describe how a vertex handles \texttt{DistanceUpdate} messages in more detail in Section \ref{distupdate}.

\textbf{Edge Deletion}: When an edge deletion is detected, the infrastructure enforces an epoch both before and after injecting the deletion. After the first epoch, the tail vertex checks whether the deleted edge was part of the shortest-path tree, specifically, whether the tail was the predecessor of the head in the shortest-path tree. If so, it sends a \texttt{SetToInfinity} message to the head to indicate that its shortest path is no longer valid, and initiates the invalidation phase.
\begin{lstlisting}[caption=Edge Deletion]
func onEdgeDeletion(u *Vertex, deletedEdge *Edge):
    onShortestPath, _ := u.SuccessorVertices[deletedEdge.Destination]
    if onShortestPath:
        sendMessage(
            deletedEdge.Destination,
            SetToInfinity
        )
\end{lstlisting}
After the invalidation phase completes, the recomputation phase begins: each invalidated vertex sends a \texttt{DistanceQuery} message to its incoming neighbors.
\subsection{Algorithmic Message Handlers}
\textbf{onDistanceUpdate}\label{distupdate} When a vertex receives a distance update, it first compares the offered distance to its current shortest known distance. If the offered distance is shorter, the vertex first notifies its current predecessor that it is no longer the predecessor. Then acknowledges the sender of the update as its new predecessor, and propagates its updated distance to all of its outgoing neighbors.
\begin{lstlisting}[caption=Distance Update Message]
// v received from u
func onDistanceUpdate(v *Vertex, u *Vertex, dist float64) {
  if v.Distance > dist:
    v.Distance = dist
    sendMessage(v.PredecessorVertex, RemoveFromSuccessor)
    v.PredecessorVertex = u
    sendMessage(u, AddToSuccessor)
    for _, edge := range v.OutEdges:
        sendMessage(
            edge.Destination, DistanceUpdate, 
            dist + edge.Weight
        )
}
\end{lstlisting}

\textbf{onAddToSuccessor}: The receiver adds the sender to its successors.
\begin{lstlisting}[caption=Add to Successor Message]
func onAddToSuccessor(v *Vertex, u *Vertex) {
  v.SuccessorVertices[u] = true
}
\end{lstlisting}

\textbf{onRemoveFromSuccessor}: The receiver removes the sender from its successors.
\begin{lstlisting}[caption=Remove From Successor Message]
func onRemoveFromSuccessor(v *Vertex, u *Vertex) {
  delete(v.SuccessorVertices, u)
}
\end{lstlisting}

\textbf{onSetToInfinity}: This message indicates that the receiving vertex's shortest path has been invalidated due to an edge deletion, and it must find an alternative path. Upon receiving this message, the vertex is marked as affected and propagates the \texttt{SetToInfinity} message to all of its successors, signaling that their shortest paths is also invalid. It then resets its predecessor and clears its set of successors.
\begin{lstlisting}[caption=Set To Infinity Message]
func onSetToInfinity(v *Vertex, u *Vertex) {
  v.Distance = inf
  v.MarkedAsInfinity = true
  for _, w := range v.SuccessorVertices:
    sendMessage(
        w,
        SetToInfinity
    )
  v.SuccessorVertices = EmptySet
  v.PredecessorVertex = EmptyVertex
}
\end{lstlisting}

\textbf{onDistanceQuery}: If the receiving vertex has a valid path from the source--i.e. it's connected, it will send the shortest path it can offer to the sender vertex.
\begin{lstlisting}[caption=Distance Query Message]
func onDistanceQuery(v *Vertex, u *Vertex) {
  if v.Distance != inf:
    edge := getEdgeBetween(v, u)
    sendMessage(
        u,
        DistanceUpdate,
        v.Distance + edge.Weight
    )
}
\end{lstlisting}
This message is sent during the recomputation phase and enables invalidated vertices to recompute their shortest paths.

\section{Evaluation\label{sec:eval}}
We evaluate our algorithm on large real-world and synthetic dynamic graphs along the following dimensions: \textit{query latency}: how quickly shortest path queries can be answered; \textit{impact of rate limiting}: how query latency is affected under rate-limited update streams; and \textit{stability}: whether the returned solutions remain consistent as the graph evolves. We also compare our approach against algorithms designed for the batch-processing model.

\subsection{Experimental Setup}

\textit{1) Implementation}: We implemented our algorithm on top of \textsc{Lollipop}~\cite{sallinen2023realtime}, a framework that supports the computational model described in \autoref{sec:compmodel}. \textsc{Lollipop} is designed for rapid prototyping of dynamic graph algorithms by emulating a distributed environment on a single machine. To process standing queries on dynamic graphs, the framework spawns a fixed number of threads. Each thread repeatedly performs two tasks: (i) dequeuing and applying topology updates while executing corresponding event handlers, and (ii) only when the topology event buffer is empty, thus prioritizing topology events, consuming messages from other vertices and executing message handlers. Threads communicate by passing messages through FIFO queues.

\textit{2) Machine}: All experiments were conducted on a commodity desktop equipped with four Intel Xeon E7-4870 v2 (Ivy Bridge) CPUs, each with 16 cores, totaling 64 cores, 1.5 TB of RAM, and an NVMe SSD. We parallelize our workload using up to 32 cores.

\textit{3) Datasets} (Table \ref{tab:dataset}): We use evolving graphs that have explicit timestamps: {hive-comments}~\cite{luo2023going} and {wikipedia-growth}~\cite{mislove2009online} and two static graphs {web-Google}~\cite{leskovec2009community} and {RMAT-20} graph~\cite{chakrabarti2004rmat}
    \footnote{The {RMAT} graph is synthetically generated using the \textit{Graph500}~\cite{graph500} benchmark parameters with a scale of 20, and assigned random edge weights in the range (0,4)},
for which we simulate randomized temporal evolution. For preprocessing, we reduce each graph to a simple graph by removing duplicate edges and assigning a weight of 1 to unweighted graphs. 

To generate the event stream with deletions, we emulate a sliding window model for topology activity. Given a window size $W$, upon observing an event with timestamp $T$, the stream probabilistically deletes events with timestamps earlier than $T - W$ using a probability parameter $\delta$. This controls the deletion event rate: $\delta = 0$ implies addition-only, while $\delta = 1$ implies a delete-heavy scenario where all edges outside the window are removed. For non-temporal datasets, we substitute timestamps with strictly increasing indices that reflect the order of edge appearance in the event stream. 

To evaluate performance across challenging scenarios, we run PageRank~\cite{page1999pagerank} on the transpose of each dataset to identify the most influential nodes in terms of inbound connectivity. We then select the top three highest-ranking vertices as source nodes for our experiments. These nodes typically result in larger and more complex shortest-path trees, making them ideal for stress testing both latency and stability under dynamic updates.

\begin{table}[h!]
\centering
\caption{Evaluated Graph Properties. Number of vertices $|V|$, number of edges $|E|$, maximum in-degree and out-degree cardinality.
}
\label{tab:dataset}
\begin{tabular}{@{\hskip 6pt}c@{\hskip 6pt}c@{\hskip 6pt}c@{\hskip 6pt}c@{\hskip 6pt}c@{\hskip 6pt}
}
\toprule
Graph & $|V|$ & $|E|$ & $\max(\text{in})$ & $\max(\text{out})$ \\ 
\midrule
{hive-comments} & 0.7M & 80M & 700K & 2.7M \\
{wikipedia-growth} & 1.9M & 40M & 200K & 7K\\
{web-Google} & 0.9M & 5.1M & 6.3K & 456\\
{RMAT(20)} & 1.0M & 16.8M & 137.8K & 42.2K\\
\bottomrule
\end{tabular}
\end{table}

\textit{4) Success Metrics}:  We evaluate three key metrics during state collection. The first is \textbf{Query Latency}, defined as the wall-clock time between issuing a state query and receiving the result. The second is \textbf{Shortest-Path Tree Stability}, which measures how much the tree structure changes between consecutive queries. We quantify this by counting the number of vertices whose predecessor in the shortest path tree differs from the previous query. The third is \textbf{Ingestion Rate}, defined as the maximum throughput of events that our system can handle.

\subsection{Baseline}
As our baseline, we adopt ReMo SSSP~\cite{sallinen2021}, which operates under the same vertex-centric, asynchronous, and shared-nothing computational model described in \autoref{sec:compmodel}. This choice is also motivated by its ability to process queries without batching constraints. However, ReMo SSSP lacks support for edge deletions, preventing its direct application to our datasets. To work around this, we dynamically construct the graph by ingesting the event stream and maintaining the evolving topology without storing vertex-level information. Upon receiving a query, we temporarily pause ingestion, run ReMo SSSP on the current graph snapshot, and collect results after convergence (i.e., once all messages have been processed).

To validate the suitability of ReMo SSSP as a baseline, we compare its performance against Galois, a well-optimized static graph processing framework. Galois is written in C++ and is recognized for its high performance in static graph analytics within shared-memory environments~\cite{nguyen2013lightweight, whang2015scalable}. Additionally, its static nature allows for extensive optimization of data layout and locality. We use Galois version 6.0 and benchmark three algorithms: \texttt{topo}, optimized for directed acyclic graphs; \texttt{AutoAlgo}, an adaptive method that dynamically selects among delta-stepping variants based on the graph’s properties; and \texttt{dijkstra}, a parallel implementation using a priority queue. Further implementation details are available in the Galois GitHub repository~\cite{galois_github}.

Each algorithm is evaluated across all datasets, with the best-performing result selected. To emulate static analysis in \textsc{Lollipop}, we cold start ReMo SSSP from scratch on the final graph once ingested. Table~\ref{tab:galois} presents a detailed runtime breakdown. Unlike Lollipop, which incrementally ingests events as a stream and processes the final topology, Galois requires the entire event log to be converted into a binary Compressed Sparse Row (CSR) format before computation. While this format significantly boosts performance through improved cache locality, it is unsuitable for dynamic graphs, any vertex or edge update would necessitate rewriting the CSR arrays. This limitation is reflected in preprocessing costs: although CSR construction is a one-time operation for static graphs, it is prohibitively expensive for evolving topologies~\cite{ediger2012stinger}.
\\
In contrast, our system ingests the graph without foreknowledge of its evolution, as described in \autoref{sec:compmodel}, which limits opportunities for memory layout or allocation optimizations. Architecturally, Galois is optimized for shared-memory systems, whereas our system is built for portability across distributed environments.
\\
Considering these fundamental differences, Galois on average incurs nearly 17× longer preprocessing and loading time compared to Lollipop’s streaming ingestion, while achieving approximately 4× faster solve times. This highlights the practicality and strength of our baseline.



\begin{table}[htbp]
\centering
\caption{Static Graph Analysis Baseline. \textsc{Galois} requires pre-processing: conversion (\textit{Conv} column in the table) from event-log to CSR format. \textsc{Lollipop} constructs the graph dynamically (\textit{Ingest}). \textit{SP (ms)} represents the average solve runtime using the top three popular vertices as source nodes.
On average across datasets, \textsc{Galois} takes approximately 50.71s for Conversion and Loading, whereas \textsc{Lollipop} completes Ingestion in just 2.97s, making \textsc{Galois} about 17× slower in this phase. For Solving, \textsc{Galois} takes around 19ms, compared to \textsc{Lollipop}'s 74ms, making it only about 4× faster in this stage. 
}
\label{tab:galois}
\begin{tabular}{@{}l@{\hskip 6pt}r@{\hskip 6pt}r@{\hskip 6pt}rr@{\hskip 6pt}r@{\hskip 6pt}r@{}}
\toprule
& \multicolumn{3}{c}{\textsc{Galois}} & \multicolumn{2}{c}{\textsc{Lollipop}} \\
\midrule
Graph & Conv(s) & Load(s) & \textbf{SP(ms)} & Ingest(s) & \textbf{SP(ms)} \\
\midrule
{RMAT(20)} & 30.97 & 0.25 & 14 & 1.73 & 34 \\
{hive-comments} & 85.73 & 0.50 & 18 & 6.65 & 126 \\
{web-Google} & 9.22 & 0.15 & 10 & 0.38 & 24 \\
{wikipedia-growth} & 75.60 & 0.43 & 33 & 3.13 & 114 \\
\bottomrule
\end{tabular}
\end{table}

\subsection{Query Latency}
This series of experiments evaluates the query latency of our proposed algorithm compared to the ReMo SSSP baseline.

\begin{figure}[h]
    \centering
    \includegraphics[width=0.4\textwidth]{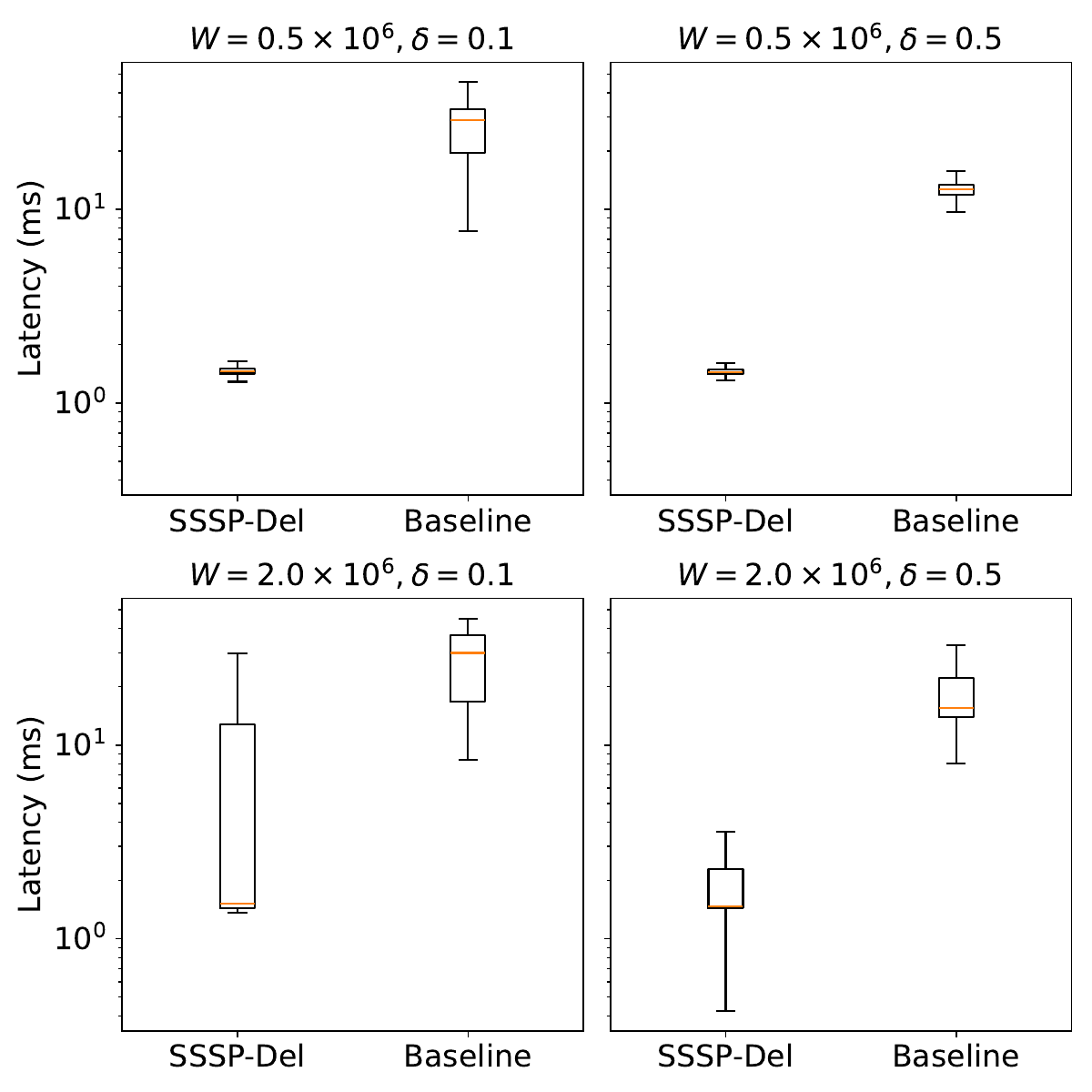}
    \caption{Comparison of query latency between SSSP-Del and the baseline, using the web-Google dataset under four different configurations: $W=\{0.5M, 2M\}\times\delta=\{0.1, 0.5\}$. The vertical axis represents the query latency in milliseconds and is in log scale.}
    \label{fig:latency-dataset}
\end{figure}

\textit{1) Dataset and Update Patterns}: We examine how different dataset configurations affect query latency, keeping the query interval and source vertex fixed. Figure~\ref{fig:latency-dataset} presents a detailed breakdown of experiments on the web-Google dataset. We evaluate four configurations using two window sizes, 0.5M and 2M edges(total dataset is $\approx$5M edges), and two delete probabilities (0.1 and 0.5). We used the most popular vertex as our source and set the query interval to 1/10 of the window size.\\
Across all four configurations, our algorithm consistently outperforms the baseline, achieving an average 14.5x speedup of the median latency. Notably, increasing the window size and delete probability has minimal impact on our system’s latency. In contrast, the ReMo SSSP baseline experiences a near 2× increase in latency when transitioning from $\delta=0.5$ to $\delta=0.1$. This is due the fact that lower deletion probability allows the graph to grow larger, and cause the latency of the baseline to increase. Similar trends were observed across other datasets, demonstrating the robustness of our algorithm to varying deletion rates and workload configurations.

\begin{figure}[h]
    \centering
    \includegraphics[width=0.4\textwidth]{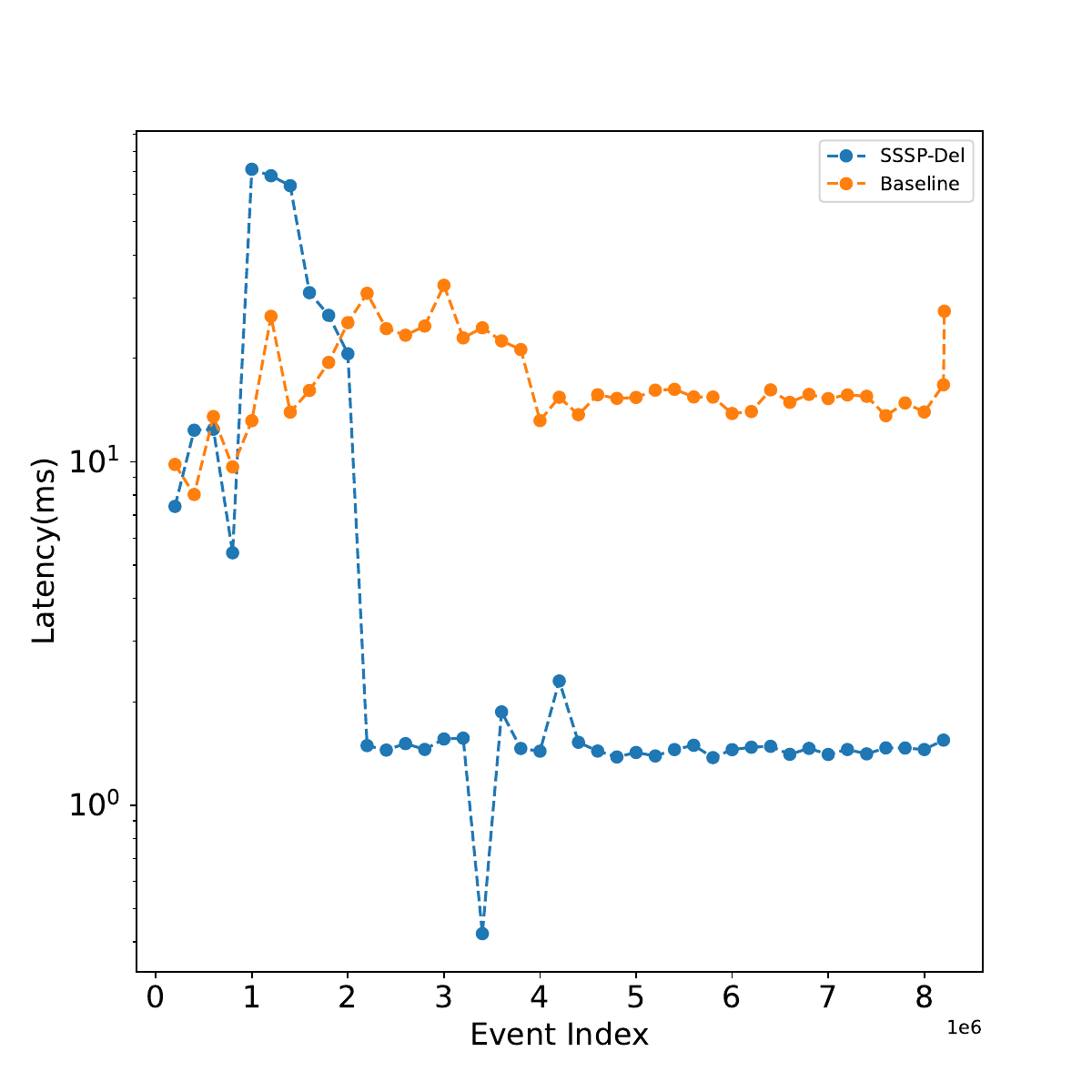}
    \caption{Comparison of query latency between SSSP-Del and the baseline, using the web-Google dataset with $W=2M$ and $\delta=0.5$. The vertical axis represents the query latency in milliseconds and is in log scale, and the horizontal axis is the event index after which the query is requested.}
    \label{fig:latency-google-grow}
\end{figure}

We also observe that our solution scales more effectively as the graph grows. In the early stages of the stream, our performance is comparable to the baseline, but as the number of ingested events increases and graph grows, our algorithm achieves up to a 40× improvement by the end of the stream. 
This trend is illustrated in Figure~\ref{fig:latency-google-grow}.

\textit{2) Source Selection}: We investigated the impact of source selection on query latency. To identify representative sources, we ran PageRank on the transpose of each full graph (i.e., reversing all edges) and selected the top three vertices with the highest PageRank scores. For this experiment, we fixed the dataset configurations and query intervals across all datasets to isolate the effect of the source choice.
We observed that the median query latency of our solution consistently outperforms the ReMo SSSP baseline across all selected sources.

\begin{figure}[h]
    \centering
    \includegraphics[width=0.4\textwidth]{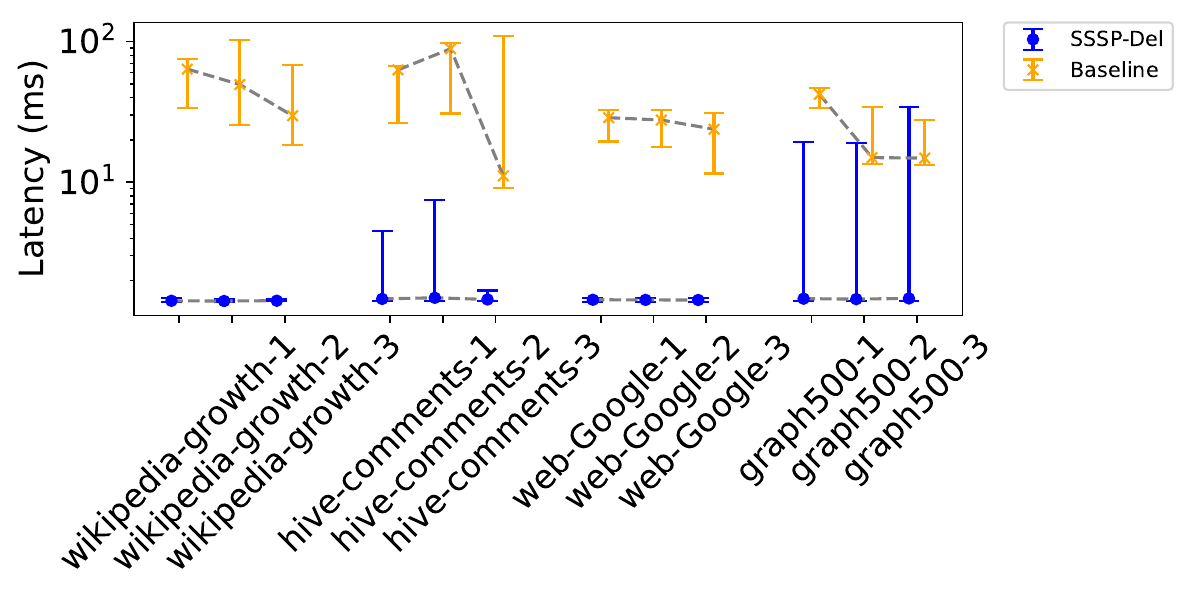}
    \caption{Latency comparison (in milliseconds, log scale) between SSSP-Del (blue) and Baseline (orange) across multiple datasets and source selections. Each x-axis label indicates a dataset and a specific source vertex (e.g., wikipedia-growth-2 refers to the Wikipedia Growth dataset with the second highest-ranked vertex as the source). Error bars represent the 25th percentile (p25), median, and 75th percentile (p75) latency across queries.}
    \label{fig:source-latency}
\end{figure}

Figure~\ref{fig:source-latency} presents the results of this evaluation. As shown, the median query latency of our proposed solution consistently outperforms the ReMo SSSP baseline across all selected sources.

\subsection{Shortest Path Tree Stability} 
\begin{figure}[h]
    \centering
    \includegraphics[width=0.4\textwidth]{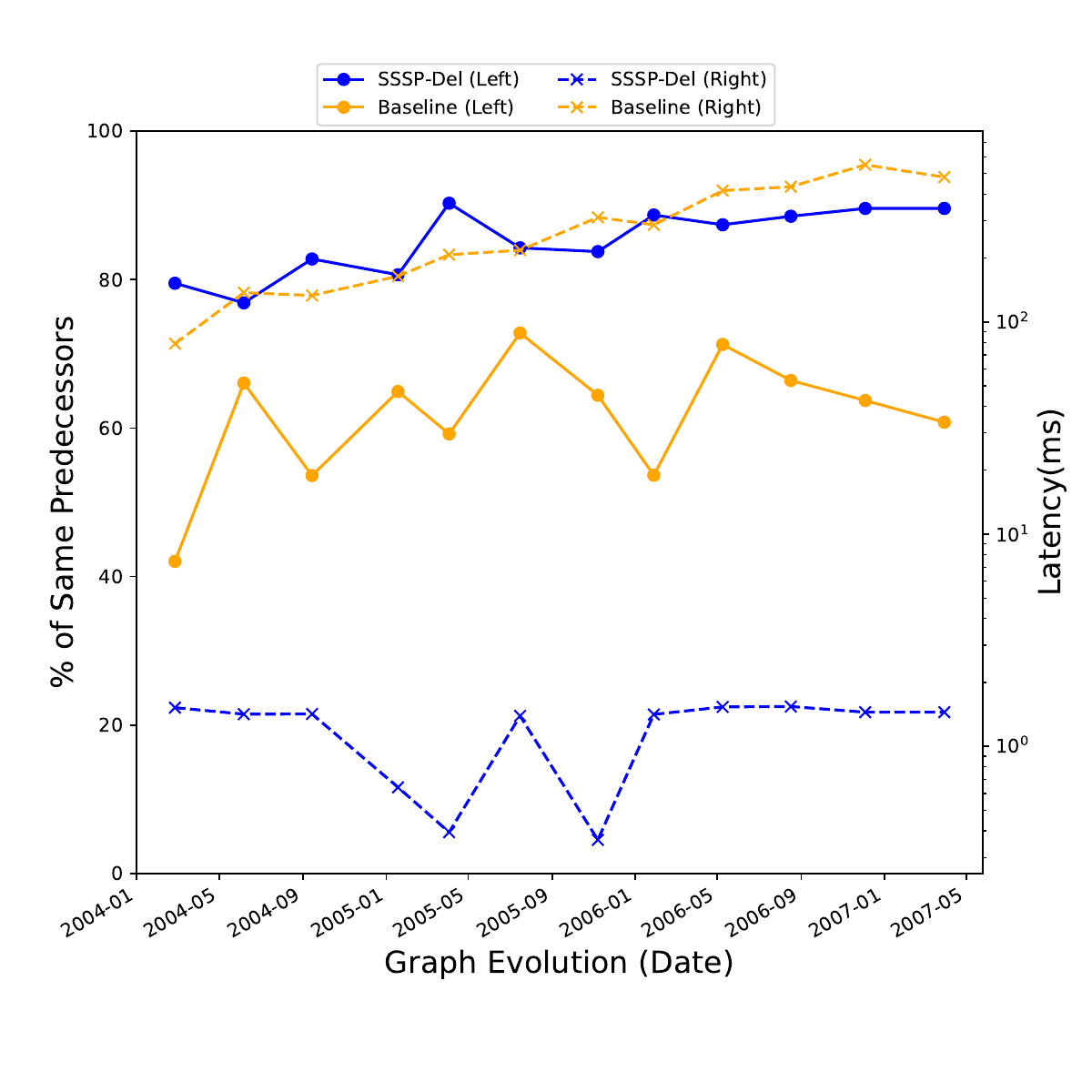}
    \caption{Stability and Result Latency over time. The graph
    evolution (x-axis) is plotted against the percentage of vertices in the current result that had the same predecessor in the previous result (y-
    axis, left) and query latency (y-axis, right). Graph: Wikipedia-growth.}
    \label{fig:stability-wikipedia-growth}
\end{figure}

An important evaluation criterion for algorithms operating on evolving graph structures is solution stability. Unlike static approaches, which recompute results from scratch for each query, our dynamic approach maintains algorithmic state as the graph evolves, allowing it to produce solutions that are incrementally updated and structurally similar to prior results. While multiple valid shortest path trees may exist for a given topology, the dynamic algorithm aims to preserve continuity by modifying only what is necessary, rather than generating an entirely new, but equally valid, solution at each step. This stability is especially desirable in real-time applications such as navigation, where small, predictable changes in the shortest path are preferred over abrupt rerouting. To evaluate this aspect, we use the Wikipedia-growth dataset as a case study. For each query, we extract the predecessor of every vertex in the shortest path tree and compute the percentage that remains unchanged compared to the previous query’s result. Figure~\ref{fig:stability-wikipedia-growth} plots both stability and query latency over time as the graph evolves, using the ReMo SSSP baseline that runs our algorithm from scratch on each snapshot. The dynamic approach demonstrates clear advantages: (i) it consistently yields more stable results across queries, and (ii) it achieves significantly lower query latency.

\subsection{Ingestion Rate}
This experiment evaluates the maximum ingestion rate our system can sustain under varying deletion probabilities. We allow the framework to ingest events at full capacity and measure the resulting ingestion rates, i.e., throughput. 

\begin{figure}[h]
    \centering
    \includegraphics[width=0.4\textwidth]{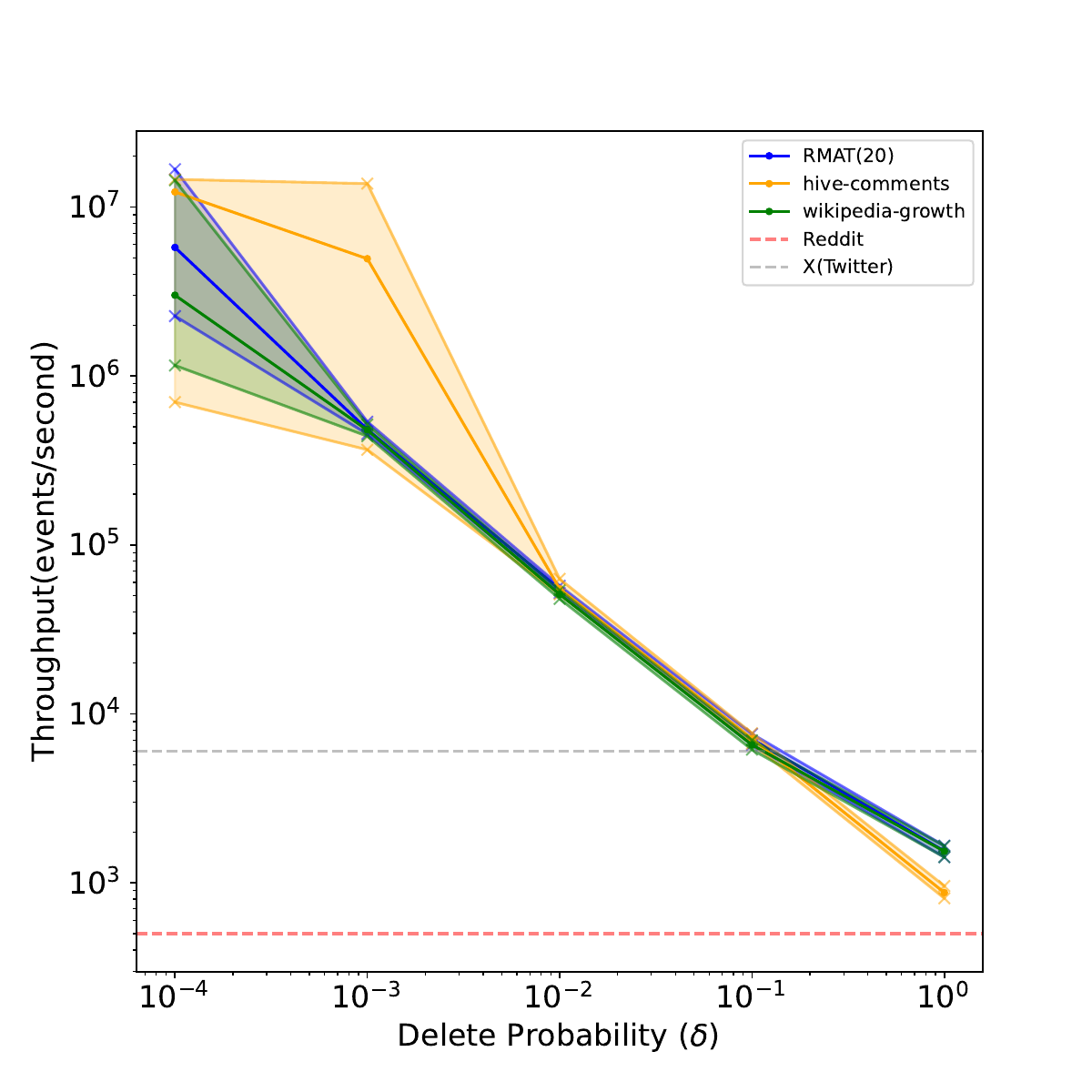}
    \caption{Throughput under different delete probabilities($\delta$). For each graph, the lines plot the p25, median, and p75 throughput achieved during graph evolution (y-axis, log scale), for different delete probabilites (x-axis, log scale). The shaded area show the range of throughputs. For comparison, the figure also presents estimated event rates of real-world systems: Reddit comments ($\approx$500 event/second) \cite{cropink2025reddit}, and Tweets posted ($\approx$6000 event/second) \cite{sayce2025tweets}.}
    \label{fig:throughput-delete}
\end{figure}

Figure~\ref{fig:throughput-delete} plots the achieved throughput across different graphs and deletion probabilities. For context, we also overlay the estimated event rates of representative real-world systems such as Reddit and X(Twitter), to highlight the practical significance of our results.

Several key observations emerge:
First, our system consistently processes events at rates that exceed those of typical real-world applications, even under high deletion rates (e.g., one deletion per addition) and when running on a commodity desktop.
Second, throughput is inversely related to the deletion probability. As expected, lower deletion rates (i.e., a higher ratio of additions to deletions) lead to fewer enforced epochs, resulting in higher throughput.

\subsection{Comparison to Related Work}
We compare our solution against GraphBolt~\cite{graphbolt2019}, a shared-memory, bulk-synchronous parallel framework designed for dynamic graph analysis. This experiment focuses on query latency under varying query frequencies during the evolution of the Wikipedia graph.

In GraphBolt, the system processes updates in fixed-size batches (defined as \#Events/\#Queries). After ingesting each batch, it applies topological updates and reconverges the algorithm to reflect the changes. We define this reconvergence time as GraphBolt's query latency.

In contrast, our asynchronous approach allows topological and algorithmic events to be processed concurrently. To ensure a fair comparison, we simulate matching query intervals by monitoring the number of ingested events: once the event streaming thread observes that a target number of events has been processed (equivalent to \#Events/\#Queries), it triggers an on-demand user query over the current graph snapshot. As previously defined, our query latency includes both the application of topological changes and the convergence time of the algorithm.

\begin{figure}[h]
    \centering
    \includegraphics[width=0.4\textwidth]{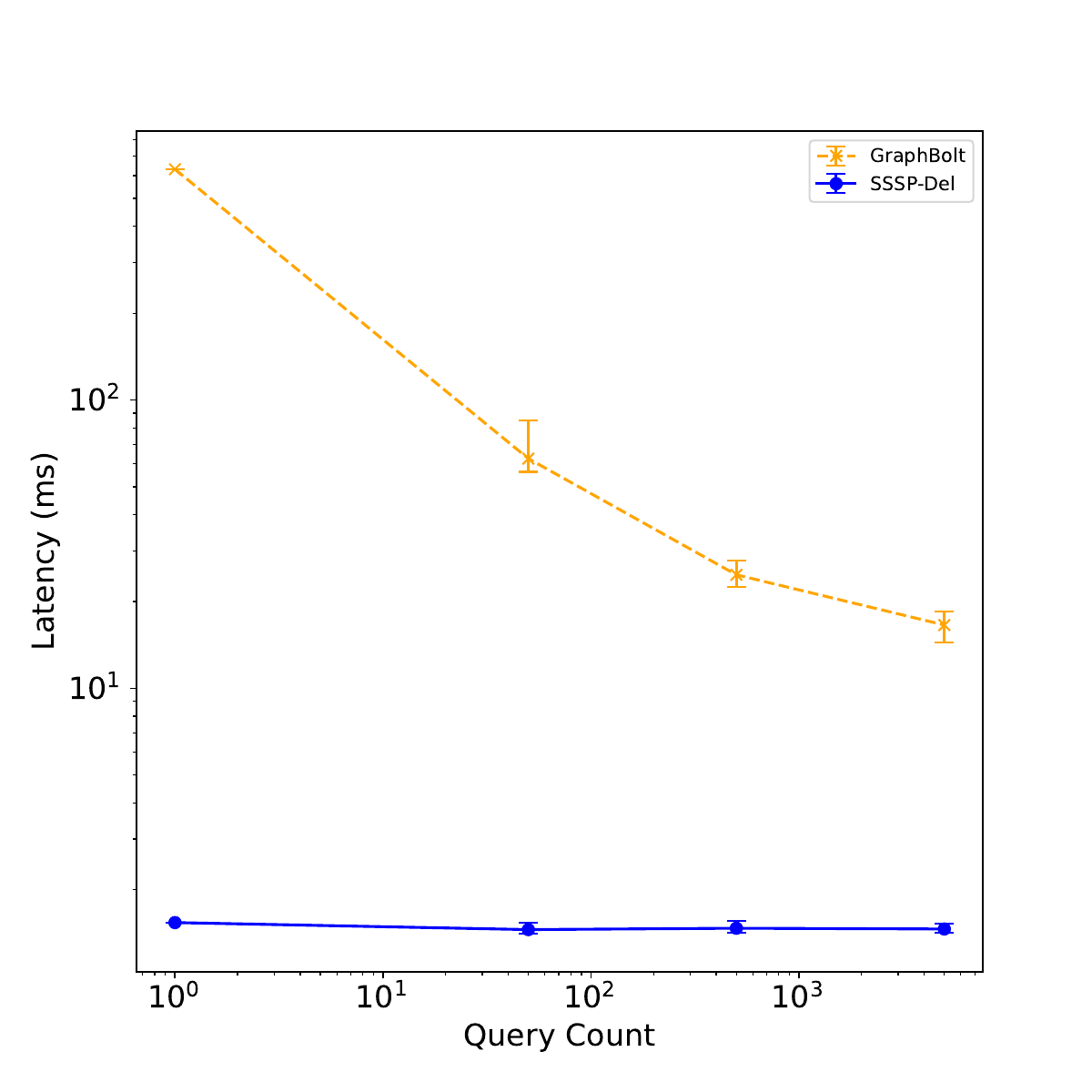}
    \caption{Comparison of Query Latency between Graphbolt~\cite{graphbolt2019} and SSSP-Del. Wikipedia-growth dataset, with deletion window size of 1000 dataset days($\approx$0.5 total dataset days), deletion probability of 0.1, and the source set to the most popular vertex determined by PageRank. For a given query
count (x-axis, log scale), we plot the p25, median, and p75 query latency in milliseconds (y-axis, log scale). The “batch” for SSSP-Del is defined as the time since the last query (presented on-demand after the event count is observed).}
    \label{fig:graphbolt}
\end{figure}

Figure~\ref{fig:graphbolt} presents the results of this evaluation. Our method consistently achieves lower query latency than GraphBolt, especially at higher query frequencies. For fine-grained queries, our system achieves up to an 11× reduction in latency, demonstrating the effectiveness of our asynchronous design in handling dynamic workloads with frequent queries.

\section{Related Work\label{sec:relatedw}}

A major challenge in solving SSSP on dynamic graphs is handling edge deletions, which can require extensive and expensive restructuring of the shortest-path tree. Consequently, some solutions, like ReMo SSSP \cite{sallinen2021}, are limited to increment-only workloads, simplifying the algorithm but reducing its generality. SSSP-Del addresses this limitation by isolating edge deletions using a "stop-the-world" strategy \cite{sallinen2019incremental} as described in Section~\ref{sec:algo}.

An alternative approach to handling decremental events in such asynchronous models involves the concept of "state generations". This technique preserves recursive and monotonic (ReMo) properties by versioning the state. When a deletion would normally break monotonicity (e.g., by increasing a shortest path distance), the affected state is advanced to a new generation. This new generational state is considered intrinsically lower than any state in the previous generation, thus maintaining a total ordering and allowing the algorithm to converge correctly \cite{sallinen2019incremental}. While this approach offers a viable alternative to the invalidation/recomputation strategy used by SSSP-Del, it may introduce its own overheads, especially in scenarios with frequent, cascading deletions.

\section{Conclusion\label{sec:conclusion}}
We present a fully dynamic algorithm for the Single-Source Shortest Path (SSSP) problem and evaluate its performance on large, real-world dynamic graphs. Designed for a vertex-centric, asynchronous, and shared-nothing computational model, the algorithm efficiently supports both edge and vertex additions and deletions. Our experimental results demonstrate that the algorithm achieves low query latency, maintains high solution stability under rapid graph evolution, and maps well to parallel and distributed platforms.

\bibliographystyle{ACM-Reference-Format}
\bibliography{reference}

\appendix
\section{Proof Of Termination and Correctness\label{app:proof}}

We aim to demonstrate two properties of the algorithm: correctness and termination. 

\subsection{Termination}
\textit{Theorem: No topological event or algorithmic message  can trigger an infinite series of messages. Therefore, the algorithm always terminates}.\\
We examine each type of message and event to show that all message chains are finite:

\begin{itemize}
    \item Passive Messages: Messages such as \texttt{AddToSuccessor} and \texttt{Remove\-From\-Successor} are local bookkeeping operations. Upon receipt, they do not trigger any further messages. 
    
    \item \texttt{DistanceUpdate} Messages: A vertex propagates a \texttt{DistanceUpdate} message only if it receives a shorter distance than its current value. Since all edge weights are strictly positive 
    a vertex can only reduce its distance a finite number of times, bounded by 1 in the worst case. Therefore, each vertex can generate and propagate this message a finite number of times.
    
    \item Edge addition Events and \texttt{DistanceQuery} Messages: Both of these may trigger a \texttt{DistanceUpdate}. As shown above, such messages can only be propagated a finite number of times. Therefore, \texttt{EdgeAddition} and \texttt{DistanceQuery} cannot lead to infinite message chains.
    
    \item \texttt{SetToInfinity} Messages: This message is used during the invalidation phase to mark vertices whose shortest paths are no longer valid. For any vertex $v$, it can send this message to each of its successors at most once, as it clears its successor list immediately after the first \texttt{SetToInfinity}, and it's not refilled during invalidation as \texttt{SetToInfinity} is the only messages in-flight during this phase. The maximum number of successors is bounded by $OutDeg(v)$, the out-degree of the vertex. Therefore, the total number of \texttt{SetToInfinity} messages in the entire graph is bounded above by the sum of out-degrees over all vertices-i.e., the number of edges—hence finite.
    
    \item Edge deletion Events: An edge deletion triggers at most one \texttt{RemoveFromIncoming} and one \texttt{SetToInfinity} message. As shown above, both of these message types lead to finite propagation. Therefore, edge deletions cannot result in infinite messaging.

\end{itemize}

Conclusion:
Every message type and topological event has bounded propagation. Since no part of the algorithm allows for infinite message generation, the system is guaranteed to converge after a finite number of steps.

\subsection{Correctness}
Let $G_t=(V_t,E_t)$ be the graph, set of vertices, and set of edges after the $t$ epochs. Between two epochs, there is either a series of edge additions, or a single edge deletion. Also, let's define $R_t$ be the directed graph formed by the successor edges $\{u, v, e_{uv}: \forall v \in \texttt{u.SuccessorVertices}, \forall u\in V_t\}$ at the end of epoch $t$. 

We claim that $R_t$ is the shortest-path tree, meaning that:
\begin{itemize}
    \item $R_t$ is a directed tree rooted at $s$.
    \item For all vertices $v$ in $R_t$, $d_t(v)=d^*(v)$ where $d_t(v)=\texttt{v.Distance}$, and $d^*(v)$ is the shortest path from source to $v$.
\end{itemize}

We will prove this by induction:\\
\textbf{Base Case}: Initially (before any changes) $R_0$ only consists of the source, and \texttt{source.Distance=0} and no edges exist.\\
\textbf{Inductive Hypothesis}: Assume at the end of epoch $t$, $R_t=\{u, v, e_{uv}: \forall v \in \texttt{u.SuccessorVertices}, \forall u\in V_t\}$ is the shortest-path tree. After ingesting the topological events between epoch $t$ and $t+1$, $R_{t+1}$ will also be the shortest-path tree.\\
\textbf{Inductive Steps}
\begin{itemize}
    \item Case 1, Only Edge Additions between $t$ and $t+1$: Suppose between epochs \(t\) and \(t+1\) a set of new edges is added, and no deletions occur.
    \begin{enumerate}
        \item No cycles introduced in $R_t$: When a new edge \((u,v)\) is added, we send a \texttt{DistanceUpdate} from $u$ to $v$.  If \(d_t(u)+w(u,v) \ge d_t(v)\), \(v\) does not change its predecessor, so \(R_t\) remains unchanged. If \(d_t(u)+w(u,v) < d_t(v)\), then \(v\) switches its predecessor to \(u\). In the old tree \(R_t\), \(v\) already had exactly one predecessor; upon switching, we remove the old successor‐edge and add \(u\to v\). This preserves exactly one parent per node and cannot create a cycle because \(u\)’s distance from $s$ in \(R_t\) was strictly less than \(v\)’s distance from $s$ (distance decreased). Hence, \(R_{t+1}\) remains a tree rooted at \(s\). The same argument holds for all the updates sent from $v$ through its outgoing edges.
        \item Optimality preserved: By the relaxation property, if the new edge yields a strictly shorter path to \(v\), then \(d_{t+1}(v)=d_t(u)+w(u,v)\), which equals the new true shortest distance \(d_{t+1}^*(v)\). That update then propagates down \(v\)’s subtree, ensuring all descendants also recompute if they improve. Monotonicity and finiteness ensure they converge to their new shortest distances. Hence, \(d_t(v)=d_{t+1}^*(v)\) for all \(v\).
    \end{enumerate}
    Therefore, when only additions occur, \(R_{t+1}\) is still a shortest‑path tree.

    \item Case 2, Single Edge Deletion between $t$ and $t+1$: Suppose between epochs \(t\) and \(t+1\) a single edge is deleted, and no other additions or deletions occur. \\
    When an edge $(u, v)$ is deleted, during the \textit{invalidation} phase, \(v\) (and recursively its successors in \(R_t\)) receives a \texttt{Set\-To\-Infinity} message, setting their distances to \(\infty\).  Then, as the system drains to convergence, each such vertex re‑computes its distance by querying all its incoming edges. Let's define the affected subtree as:  
\[
T \;=\;\{\,v\}\;\cup\;Succ_t(v)\;\cup\;Succ_t\bigl(Succ_t(v)\bigr)\;\cup\;\dots
\]  
That is, all vertices whose unique path in \(R_t\) passed through the deleted edge \((u,v)\).

\begin{itemize}
    \item Vertices Outside the Affected Subtree: If \(x\notin T\), then its path in \(R_t\) did not use the edge \((u,v)\).  Therefore, the deletion does not trigger a \texttt{Set\-To\-Infinity} at $x$, and therefore the vertex and its predecessors will not be marked, so $d_t(x)=d_{t+1}(x)=d^*_{t+1}(x)$.
    \item Vertices Inside the Affected Subtree: We now focus on vertices \(x\in T\).  By construction, at the moment just after the invalidation phase, each such \(x\) has been received a \texttt{SetToInfinity} (directly or via propagation), so \(\texttt{v.Distance}=\infty\). During the recomputation phase, they then issue \texttt{DistanceQuery} messages to all their incoming neighbors.
    
We will prove by induction that every vertex $x$ in $T$ eventually computes its correct new shortest-path distance, $d^*_{t+1}(x)$. Let $x_1, x_2, ..., x_m$ be the vertices of the affected subtree $T$, ordered such that $d^*_{t+1}(x_1)\leq d^*_{t+1}(x_2) \leq ... \leq d^*_{t+1}(x_m)$.\\
\textbf{Claim}: The algorithm correctly computes $d^*_{t+1}(x_i)$ for all $i$ from 1 to $m$.
\begin{itemize}
    \item \textbf{Base Case}, $i=1$: Consider $x_1$, the vertex in $T$ with the smallest final shortest-path distance. Let $p$ be the predecessor of $x_1$ on its new shortest path. So, $d^*_{t+1}(x_1) = d^*_{t+1}(p) + w(p, x_1)$. By the properties of shortest paths, $d^*_{t+1}(p) < d^*_{t+1}(x_1)$. This implies that the predecessor $p$ cannot be in the affected subtree $T$. If $p$ were in $T$, it would have a shortest-path distance smaller than $x_1$, which contradicts our definition of $x_1$ as the vertex in $T$ with the minimum new distance. Therefore, $p$ must be a vertex outside of $T$. The deletion did not affect $p$'s distance, so its distance is already correct. When $x_1$ sends a \texttt{DistanceQuery} to $p$, $p$ will respond with its correct distance, $x_1$ will receive this valid path offer and by the relaxation property, will correctly compute its new shortest-path distance $d^*_{t+1}(x_1)$.
    \item \textbf{Inductive Step}: Assume that the first $k-1$ vertices in the sorted list, $x_1, x_2, ..., x_{k-1}$ have all eventually converged to their correct new shortest-path distances. Now, consider vertex $x_k$. Let p be its predecessor on its new shortest path in $G_{t+1}$. We know that $d^*_{t+1}(p) < d^*_{t+1}(x_k)$. The predecessor $p$ can be in one of two categories:
    \begin{itemize}
        \item $p$ is outside the affected subtree $T$. As in the base case, $p$'s distance is already correct. It will respond to $x_k$'s \texttt{DistanceQuery} with the correct path information.
        \item $p$ is inside the affected subtree $T$. Since $d^*_{t+1}(p) < d^*_{t+1}(x_k)$, $p$ must be one of the vertices in the set $x_1, x_2, ..., x_{k-1}$. By our inductive hypothesis, $p$ is guaranteed to eventually converge to its correct new shortest distance $d^*_{t+1}(p)$.
    \end{itemize}
    When $x_k$ sends its \texttt{DistanceQuery} to $p$, $p$ may not have finished its own recomputation and might not yet know its final, optimal distance. However, this does not break the logic. Once $p$ does converge to its correct distance $d^*_{t+1}(p)$, the algorithm ensures that it will propagate this new, shorter distance to all of its outgoing neighbors by sending \texttt{DistanceUpdate} messages . Therefore, $x_k$ is guaranteed to eventually receive the correct path information from its true predecessor $p$, even if it's not in immediate response to its initial query.
\end{itemize}

\end{itemize}
\end{itemize}

We conclude that at the end of epoch $t+1$, $R_{t+1}$ is once again the shortest-path tree of the updated graph.

\end{document}